\documentclass[pre,aps,floatfix,superscriptaddress,two column]{revtex4-2}
\usepackage{graphicx}
\usepackage{dcolumn}
\usepackage{latexsym}
\usepackage{hyperref}
\usepackage{amsmath, amsthm, amssymb}
\usepackage{epsfig}
\usepackage{bm}
\usepackage{geometry}
\usepackage[dvipsnames]{xcolor}
\usepackage{placeins}
\usepackage[normalem]{ulem}
\begin{document}


\title{Spontaneous Rotation of a Symmetric Inclusion in Chiral Active Bath}

\author{Abhra Puitandy}
\email{abhrapuitandy.rs.phy22@itbhu.ac.in}
\affiliation{Department of Physics,Indian Institute of Technology(BHU), Varanasi-221005,India}

\author{Shradha Mishra}
\email{smishra.phy@iitbhu.ac.in}
\affiliation{Department of Physics,Indian Institute of Technology(BHU), Varanasi-221005,India}

\date{\today}

\begin{abstract}
We study the dynamics of a circular passive inclusion, termed a torquer, in a bath of chiral active Brownian particles. Despite being geometrically symmetric and non-motile, the torquer exhibits persistent rotation due to spatially inhomogeneous torques arising from angularly biased collisions with active particles. This interaction-driven symmetry breaking does not rely on shape anisotropy or external forcing. Through simulations, we identify two distinct regimes of rotation: one dominated by density gradients at low chirality, and another by increased impact frequency at high chirality. Our results highlight how nonequilibrium interactions in chiral active media can induce motion in symmetric objects, offering a new perspective on symmetry breaking in active systems.
\end{abstract}

\maketitle

\section{Introduction\label{secI}}
Active matter refers to systems composed of self-driven constituents that convert internal or environmental energy into persistent motion, thereby violating detailed balance at the microscopic scale. These systems span a wide range of scales, from cytoskeletal 
filaments\cite{mahmud2009directing} and bacterial colonies\cite{fodor2018stat} to animal herds\cite{hueschen2023wildebeest} and human crowds\cite{bottinelli2016emergent}, and exhibit rich emergent collective phenomena\cite{cates2010arrested,dikshit2023ordering,fily2012athermal,redner2013structure,stenhammar2013continuum,thompson2011lattice,wysocki2014cooperative} such as swarming\cite{chate2020dry,vicsek1995novel,liebchen2017collective}, phase separation\cite{bialke2015active}, dynamic clustering\cite{gokhale2022dynamic},and pattern formation\cite{liebchen2017collective}. A particularly rich class of active systems is composed of chiral active particles, which, in addition to self-propulsion, exhibit a persistent angular drift due to either internal asymmetry or external torques. These chiral particles have been observed in natural systems such as sperm cells and bacteria 
 \cite{jennings1901significance}, as well as in synthetic colloidal systems, and are known to exhibit diverse phase behaviour\cite{semwal2024macro,wang2024condensation} and  nonequilibrium phenomena including hyperuniform phases \cite{lei2019nonequilibrium,huang2021circular,zhang2021passive,kuroda2023microscopic}, self-reverting vortices \cite{caprini2024self}, caging in active glasses \cite{debets2023glassy}, and demixing in binary mixtures \cite{ai2023spontaneous,reichhardt2019reversibility,kushwaha2024chirality}.\\
An important question in chiral active systems is how they interact with foreign inclusions such as walls, obstacles, or passive particles. Several studies have shown that chiral active particles can exert net forces and torques on asymmetric inclusions, leading to effects such as persistent rotation of microgears \cite{li2023chirality} in bacterial suspensions \cite{sokolov2010swimming} or mobile inclusions in active nematics \cite{ray2023rectified}. These interactions can result in non-reciprocal momentum exchange and symmetry breaking at the level of the embedded object, even in the absence of external drive.\\
Prior studies typically relied on geometric asymmetry or mobile defects to break symmetry and induce motion \cite{shankar2019hydrodynamics}.
However, the role of geometrically symmetric foreign objects in chiral active systems remains less explored. Specifically, it is not yet well understood whether a circular object—one that is spherically symmetric—can exhibit persistent rotation solely due to the structure of its interaction with the active bath.\\

In this work, we study a circular inclusion immersed in a bath of chiral active Brownian particles (ABPs) \cite{bechinger2016active}. This object, which we refer to as a \textit{torquer}, is not self-propelled, but dynamically reorients in response to spatially inhomogeneous torques arising from ABP collision. The name reflects its key role as a torque-responding body, distinct from both passive inclusions and active agents.\\
Crucially, while the torquer is geometrically isotropic, the torque imparted by an active particle depends on the angular location of impact of active particles. This leads to a non-uniform torque distribution around the object's boundary, effectively breaking angular symmetry without invoking any shape anisotropy. In contrast to previous models that rely on geometric or motile asymmetry, here the symmetry-breaking mechanism is encoded in the local interaction and chirality of the bath.\\
Our comprehensive numerical study reveals that this angular inhomogeneity is sufficient to induce robust and persistent rotation of the torquer. We identify two distinct rotational regimes: (i) At low chirality, the rotation is driven by density inhomogeneities of active particles at the surface of the torquer, which exert uneven torque around its perimeter; (ii) At high chirality, it is dominated by more frequent impact of ABPs at its perimeter.\\
Recent works with surface \cite{pohl2014dynamic,eswaran2024synchronized}, mass \cite{olarte2018thermophoretic}, or shape asymmetries \cite{aubret2018targeted} have demonstrated diverse mechanisms of induced motion in colloids, including optothermal propulsion \cite{chand2025optothermally} and emergent edge flows in spinner assemblies \cite{van2016spatiotemporal}. These studies underscore how asymmetry—whether structural or interaction-driven—can serve as a powerful design principle in active matter systems. Our work complements these efforts by showing that persistent rotation can emerge even in symmetric inclusions via angular interaction rules.\\
The rest of the paper is organized as follows: Section II  describes the model details. Section III, discusses the detailed results of the system described in section II. Finally, in Section IV, we summarize our main findings and discuss their broader implications for 
understanding 
symmetry breaking and        
induced motion in chiral active systems.\\
\begin{figure*}[hbt]
 \centering    
 \includegraphics[width=0.89\textwidth]{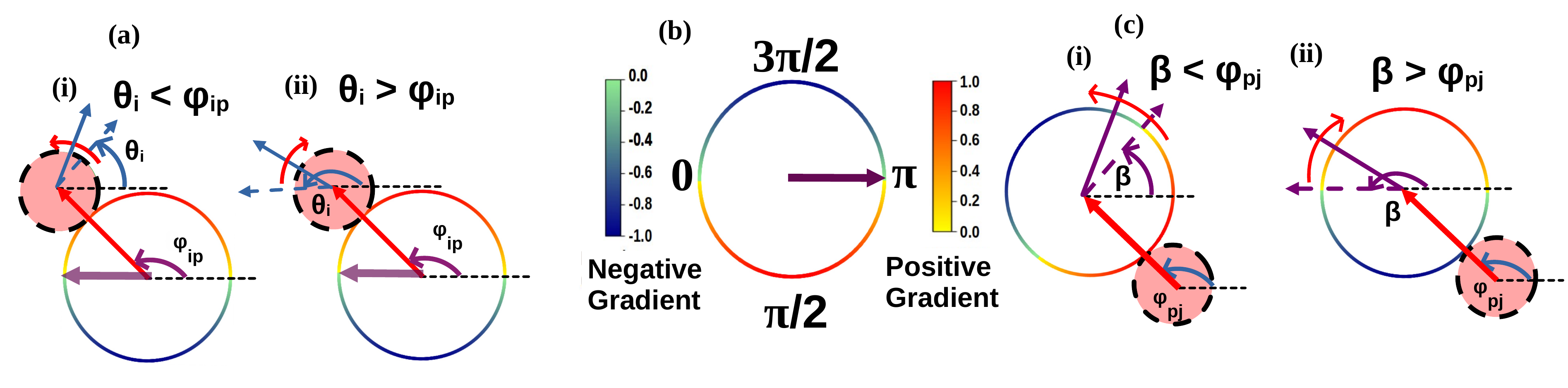}
 \caption{\textbf{Thematic representation of the rotational dynamics of the active particles and inclusion}: Fig.(a) 
 The blue (dotted) solid arrows pointing radially outwards from the active particle center denote the orientation (before) after the interaction with the inclusion, with $\theta_i$ being the angle made by it with the reference direction. The red curved arrow over the active particle denotes the direction of its reorientation after interacting with the inclusion. The faint magenta arrow in the inclusion denotes the orientation of inclusion. The straight red arrow denotes the direction of $\mathbf{r}_{ip}$. In fig.(b) the magenta arrow denotes $\beta$ which is 0 in this case. The positive and negative gradient shows the magnitude of the torque term (both of which attain a peak value perpendicular to $\beta$)  and its representative color from the color bar which denotes the magnitude of the torque acting at that coloured position on the periphery of the inclusion. The continuous gradient shows the continuous change in their magnitude over the perimeter of the inclusion. Four possible values of  $\phi_{pj}$ are labelled  which denotes the angle made by the position vector pointing from the active particle center to the inclusion’s center at the labelled respective positions. 
 Fig. (c) denotes the mechanism of rotation of the inclusion. The straight red arrow denotes the direction of $\mathbf{r}_{pj}$. The maroon (dotted) solid arrows pointing radially outwards from the particle center denote the $\beta$ (before) after the interaction with the active particle in its contact. The curved red arrow denotes the direction of rotation of the inclusion after interacting with the active particle.}
 \label{fig.1}
 \end{figure*}
\section{Model}\label{sec:mod}
We consider a two-dimensional system consisting of a large, immobile inclusion of radius $a_p$ placed at the center of a square box of side length \textit{L}, surrounded by a bath of chiral active Brownian particles (ABPs) which will be referred to as active particles in the rest of the manuscript. Each active particle is modeled as a disk of radius $a_{ac}$, and the system evolves under overdamped dynamics in the absence of thermal translational noise. 
Each active particle has a position $\mathbf{r}_i$ and an orientation angle $\theta_i$, which defines its self-propulsion direction $\mathbf{n}_i=(cos\theta_i,sin\theta_i)$. The dynamics of the $i^\text{th}$ active particle are governed by the following equations of motion:
\begin{equation}\label{eq:1}
\frac{d\mathbf{r}_i}{dt} = v_0\mathbf{n}_i+\mu\sum_j \mathbf{F}_{ij}\\\
\end{equation}
\begin{equation}\label{eq:2}
\partial_t\theta_i = -\gamma\sin(\theta_i-\phi_{ip})+\sqrt{2D_r}\eta_{i}+\omega
\end{equation}
Here, $v_0$ is the self-propulsion speed, $\mu=1$ is the mobility and $D_r$ is the rotational diffusion constant which sets the timescale in the active particles to change their orientation. An active particle experiences two types of interaction: one due to mutual exclusion with soft repulsion from active particles and with inclusion (when in contact). The  soft repulsive force $\mathbf{F}_{ij}$ exerted on particle \textit{i} by particle \textit{j}, is given by $\mathbf{F}_{ij}=F_{ij}\hat{\mathbf{r}}_{ij}$ where
\begin{equation*}
F_{ij}=
\begin{cases}
k(a_i+a_j-r_{ij}), & \text{if } (r_{ij} < a_i+a_j) \\
0, & \text{otherwise } 
\end{cases}
\end{equation*}
where $r_{ij} =|\mathbf{r}_i - \mathbf{r}_j|$, and $k$ is the repulsion stiffness. The second type of interaction an active particle experiences is due to the torque at the surface of the inclusion.  This interaction tries to effectively reorient the active particles away from the inclusion's surface. 
The $\phi_{ip} = arg(\mathbf{r}_i - \mathbf{r}_p)$.  The cartoon of this interaction term is shown in Fig. \ref{fig.1} (a), depicting the two cases (i) and (ii) of incoming direction of orientation of active particle $\theta_i$. The solid and dashed circles represent the inclusion and active particles respectively. The solid and dashed arrow in the body of active particle represents the incoming and outgoing direction of orientation respectively.  $D_r$ is the rotational diffusion constant, $\eta_i(t)$ is Gaussian white noise with zero mean and unit variance, and $\omega$ is the intrinsic chirality of the active particle. \\
Now we turn to the dynamics of the inclusion, which is also address as torquer's at some places. The inclusion's position is fixed but it is allowed to rotate about its center. 
We assign with inclusion an internal asymmetry  in the plane of the inclusion, which lead to the inhomogeneous response of active particle at the perimeter of it. The direction of asymmetry is marked with magenta solid arrow in the cartoon shown in Fig. \ref{fig.1} (b) and is denoted as $\beta(t)$. The dynamics of $\beta(t)$ can be given by 
\begin{equation}\label{eq:3}
\frac{d\beta}{dt}=-\gamma\sum_j \sin(\beta-\phi_{pj})\\
\end{equation}
The angles marked at the four corners on the inclusion are the four typical approaching directions of active particle $\phi_{pj} = arg(\mathbf{r}_p - \mathbf{r}_j)$. The colors in the color bar represents the corresponding contributions of torque generated to inclusion by active particle at those positions. 
\begin{figure*} [hbt]
 \centering    
\includegraphics[width=0.85\textwidth]{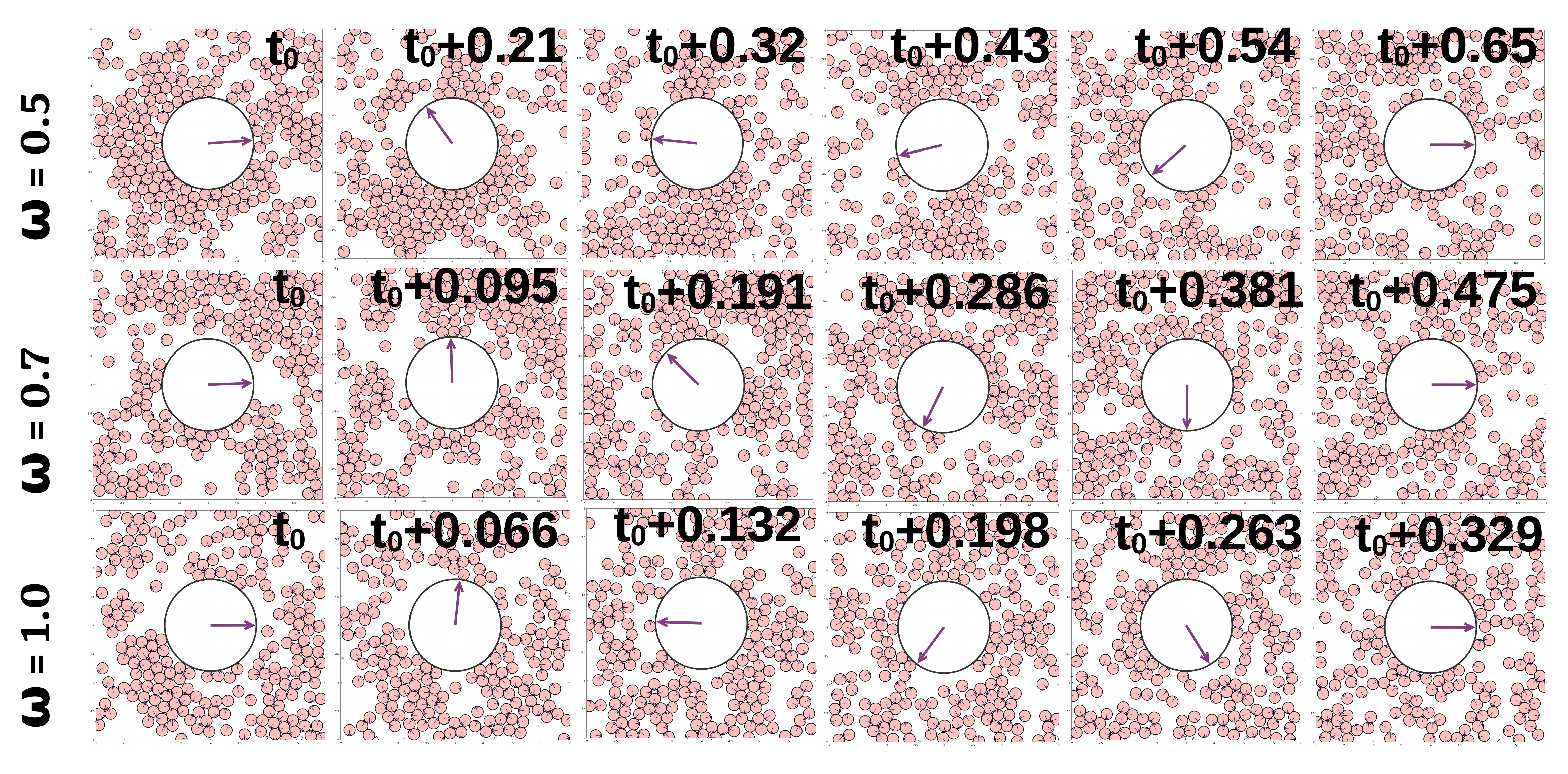}
 \caption{The active particles are denoted by the smaller solid circles.The inclusion is depicted by the larger circle at the centre wherein the solid magenta arrow denotes the inclusion orientation after interaction with all the active particles distributed as shown in the respective snapshots. The figure shows a comparison of the time taken to execute a full rotation for a single ensemble for size ratio ($S=8$) having different chirality values, $\omega$: 0.5,0.7,1.0. In each of the rows ,$t_0$ denotes the reference times from when a single rotation is measured.}
 \label{fig.2}
 \end{figure*}
 A schematic representation of the variation of torque over the inclusion circumference is shown in Fig. \ref{fig.1} (c) for the two cases (i) $\beta < \phi_{pj}$ and (ii) $\beta > \phi_{pj}$. In both the cases the result of torque term in Eq. \ref{eq:3} tries to reorient $\beta(t)$ in the direction of $\phi_{pj}$.\\
 
The above three equations \ref{eq:1} to \ref{eq:3} are numerically integrated with Euler integration scheme for time. The initial condition we started with random homogeneous non-overlapping initial positions and orientations of active particles on the two-dimensional substrate of size $L = 80 a_{ac}$.  Periodic boundary conditions are applied in both directions of the square box. 

 We define the smallest length, the size of the active particle $a_{ac}$
and the time scale in our system as $\tau=D_R^{-1}$. All other lengths and times are measured in units of $a_{ac}$ and $\tau$ respectively.

 The area fraction of active particles, defined as $\phi_a=N_a\pi a_{ac}^2/L^2$ is fixed at $0.5$. The strength of  chirality $\omega$ and size ratio $S = \frac{a_p}{a_{ac}}$ are the two control parameters in our simulations.  The $\omega$ is varied in the range $[0, 1.5]$ and $S$ in the range $[6, 16]$.
 The self-propulsion speed of active particles $v_0 = 1$ and hence the typical persistent length of active particles $l = v_0/D_r ~ 10$, hence we considered the size of inclusion either of the order or larger than $l$. The rotational diffusion constant $D_r$ is fixed to $0.1$ or $\tau = 10$. The small time step
of integration $dt = 10^{-4} \tau$. The above three equations are simulation for total number of time steps $T = 3 \times 10^5 = 30 \tau$.  A single simulation step is counted once all the active particles positions and orientations are updated once. The observations are performed after $10 \tau$, when the steady state is reached. The steady state is characterized when we do not see any statistical pattern in the dynamics of the particles. The averaging is performed over total $20 \tau$ times in the steady state and $80$ to $150$ independent initial realizations.   The number of active particles is $1000$ across all parameter sets. 

\begin{figure*}[htb]
 \centering    
 \includegraphics[width=0.58\textwidth]{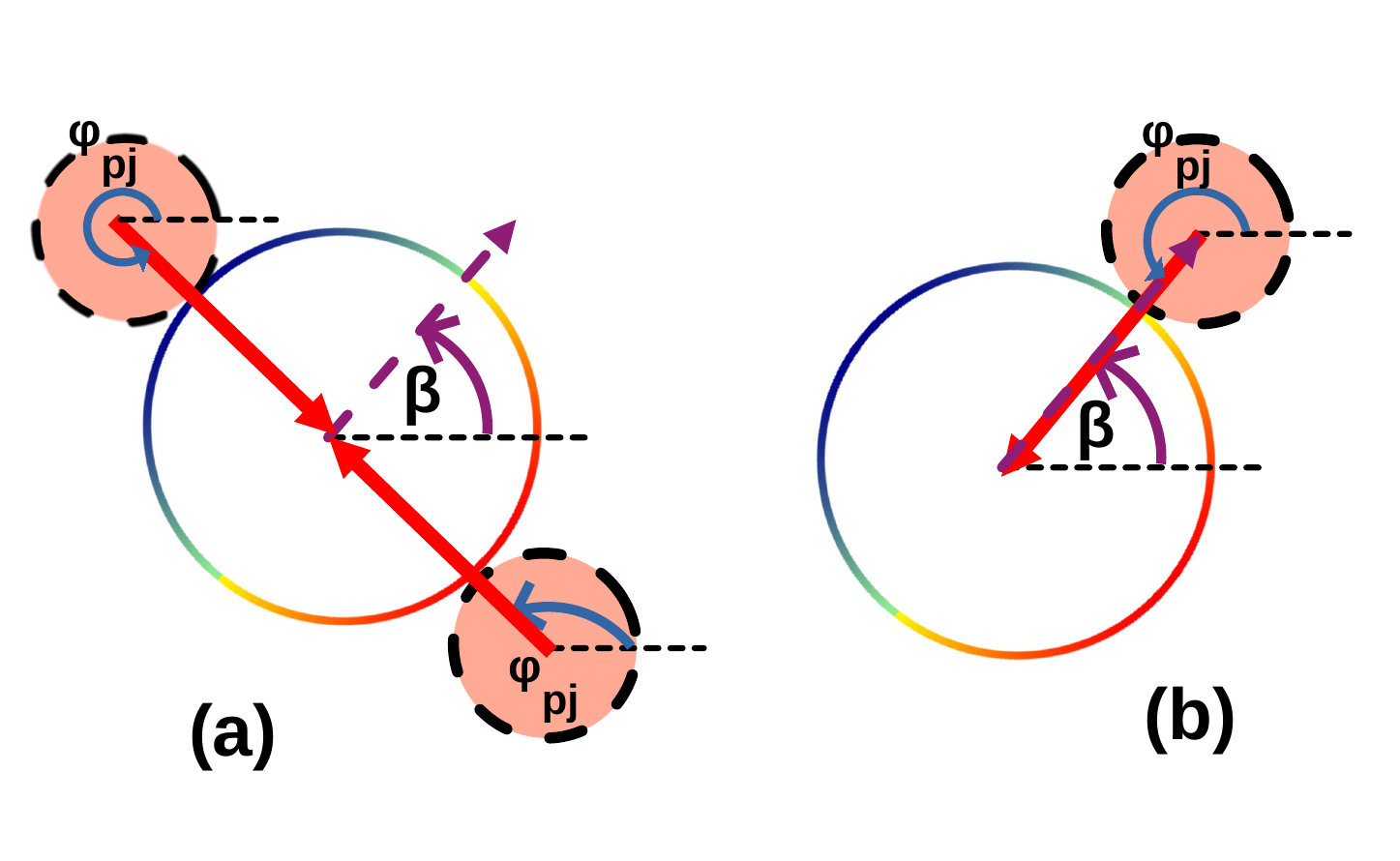}
 \caption{\textbf{Condition for no rotation of the torquer}: The solid dotted circle denote the active particle and the larger circle denote the inclusion.The solid red arrow denote the direction of $\mathbf{r}_{pj}$.The dotted magenta arrow denote the inclusion orientation before the interaction. (a) diametrically placed active particle (b) $\phi_{pj}$ coincides with $\beta$. This result corresponds to a condition in which no net torque builds up over time, and the inclusion remains statistically stationary in its angular coordinate.}
 \label{fig.3}
 \end{figure*}

\section{Results}\label{sec:res}

We now present the rotational dynamics of inclusion, by varying two key parameters—the chirality $\omega$ of the active Brownian particles and the size ratio $S=a_p/a_{ac}$. \\
Fig. \ref{fig.2} illustrates the time evolution of the inclusion's orientation $\beta(t)$ for different chirality values for a particular size ratio $(S = 8)$. The columns from left to right are at different times. With time the direction of inclusion reorients in an anticlockwise manner.  In each row, the total time is chosen such that the inclusion completes one full $2 \times \pi$ rotation. The time to complete one full rotation increases with decreasing chirality. The animations corresponding to these snapshots for three different chirality are shown in \ref{sec:appndx} \\
The magnitude of chirality introduces a natural length scale: the radius of curvature of the active particle's trajectory, given by $R_c = v/\omega$. When the size ratio $S$ is larger than $R_c$, the inclusion acts as an obstacle with respect to the curved paths of the active particles. Conversely, when $R_c > S$, the active particles follow nearly straight paths, making them more sensitive to the inclusion. Hence the relative ratio of $R_c$ $vs$. $S$ differentiates the low and high chirality regimes for the system. \\
At low chirality, the angular trajectory grows slowly and exhibits stronger fluctuations, whereas at high chirality, the rotation proceeds more quickly and smoothly. This indicates that the net torque experienced by the inclusion is present throughout, but becomes more directionally consistent at higher $\omega$, leading to an increase in angular velocity. \\
 To understand the details of the dynamics of inclusion, it is useful to examine the structure of the torque acting on the inclusion and how it is influenced by the spatial arrangement and motion of the surrounding active particles. The torque is governed by the angular difference between  $\beta$ and the angle $\phi_{pj}$. As such, the torque term in the angular dynamic works as a collision avoidance \cite{chepizhko2013optimal,gregoire2004onset} mechanism, deflecting particles away from potential overlap. This opens up two possible scenarios (illustrated in Fig. \ref{fig.3}) for the inclusion's orientation $\beta$ to remain unaffected. For a single inclusion-active particle interaction, the torque term is ineffective if the inclusion's orientation 
 and $\mathbf{r}_{pj}$ are either parallel or antiparallel to each other. But, when many active particles surround the inclusion, any diametrically opposite arrangement of active particle pairs does not account for any net angular displacement to the inclusion as the individual torque contribution from their contacts would nullify each other. The two such possible scenarios are shown in a schematic diagram in Fig. \ref{fig.3}(a-b). In the next paragraph we will discuss how the inhomogeneous distribution of active particles at the perimeter of the inclusion affects its rotation.\\
 
 The snapshots in Fig.\ref{fig.2} as well as the animations \ref{sec:appndx} show that the distribution of active particles at the surface of the inclusion is visibly inhomogeneous. 
In this section we study how the orientation of the inclusion depends on the distribution of the active particles at its surface. We compute a time-averaged cross-correlation between the inclusion's orientation $\beta$ and the  $\psi_{\text{gap}}$, where $\psi_{\text{gap}}$ is the angular location with least population of active particle at the surface of the inclusion. We calculate $\psi_{\text{gap}}$ in the following manner: the region around the inclusion is divided into angular sections ($\Delta\theta$) where, $\Delta\theta = 2 \arcsin{(\frac{1}{1+S})}$ and the number of sectors is $2\pi/\Delta\theta$.
 
\begin{figure*}[htb]
 \centering    
 \includegraphics[width=0.97\textwidth]{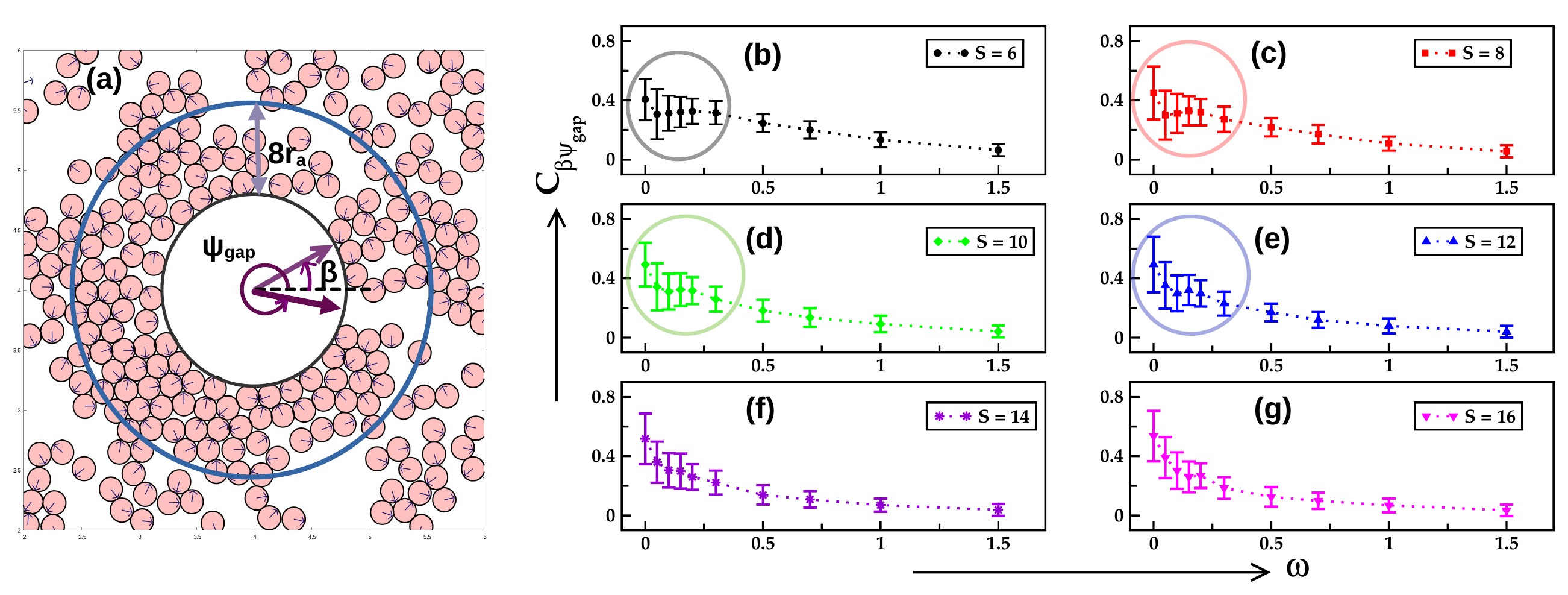}
 \caption{In Fig.(a) the blue circle denotes the periphery of the region which is divided into sectors each of angular width $\Delta\theta$. The number of active particles (denoted by smaller solid circles) in each of the sectors surrounding the inclusion periphery is then counted. The gap angle is denoted by $\psi_{gap}$ and it denotes the least occupied (by active particles) sector. Fig.(b)-(g) denotes time-averaged cross-correlation $\langle \cos(\beta - \psi_{\text{gap}}) \rangle$ between the inclusion’s angular position and the location of the largest angular gap in active particle distribution as a function of chirality $\omega$ and  for different size ratios $S = 6, 8, 10, 12, 14, 16$.} 
 \label{fig.4}
 \end{figure*}

   The number of such sectors increases with the inclusion size ratio $S$, providing finer angular resolution. At each time step, a circular shell of radius $r_{\text{shell}}=a_p+8a_{ac}$ is considered as shown in Fig. \ref{fig.4} (a), and the number of particles within each sector is recorded and averaged 
   over  $100$ independent realisations. The sector with the lowest occupancy is identified, and its central angle is defined as $\psi_{\text{gap}}$. The directional correlation between the inclusion and this depleted sector is measured via the cross-correlation function:
 \begin{equation*}
C_{\beta\psi_{gap}} = \left\langle cos{(\beta-\psi_{gap})} \right\rangle_t.
\end{equation*}%
 
 Fig. \ref{fig.4} (b-g) shows how $C_{\beta\psi_{gap}}$ varies with chirality $\omega$ for different inclusion's sizes. $C_{\beta\psi_{gap}}$ is maximum for an achiral system and a pronounced decrease is observed as chirality increases from zero to small finite values. At $\omega = 0$, active particles move without intrinsic curvature, and the angular dynamics of the inclusion are slow. In this regime, both the inclusion’s orientation and the $\psi_{gap}$  evolve slowly with time. As a result, the relative angle between them remains more stable across time steps, leading to a higher time-averaged correlation. When chirality is introduced, even weak curvature in particle trajectories causes them to approach the inclusion from a wider range of directions, shifting the position of the depleted sector more frequently. This change breaks the previous alignment and results in a  drop in the measured correlation as $\omega$ increases from zero.\\
As chirality increases further, which is shown in marked circles in Figs. \ref{fig.4} (b-e), the correlation rises and reaches a peak at intermediate $\omega$. In this range, particle motion becomes sufficiently curved and persistent to induce more consistent patterns in where particles accumulate and where gaps tend to form. The least-occupied region appears more regularly at specific angular locations relative to the inclusion, improving the average alignment and increasing the correlation.\\
At higher chirality, the correlation gradually decreases again. Particles now follow tight curved paths and interact with the inclusion more frequently, but with shorter contact durations. This leads to more uniform exploration of the inclusion boundary and suppresses persistent low-density regions. Consequently, the angular position of the gap becomes more evenly distributed around the inclusion's surface, and the average alignment with the inclusion orientation weakens. 
\\
This trend also depends on the inclusion size ratio. For smaller inclusions (Fig. \ref{fig.4} (b-e)), the features become prominent: the initial drop, the peak, and the high-$\omega$ decay are more pronounced. As S increases, the number of angular sectors grows, and the contribution of any single depleted sector to the overall correlation is reduced. The larger perimeter also supports a greater number of particle interactions, which smooths out density variations. As a result, the correlation curve becomes flatter, and for the larger inclusions (Fig. \ref{fig.4} (f,g)), the peak becomes less distinct.\\
To quantify the rotational dynamics of inclusion we calculate the angle auto-correlation of $\beta$(t) defined by 
\begin{equation*}C(t) = \langle \cos[\beta(t + \tau) - \beta(t)] \rangle_t
\label{acf}
\end{equation*}
Here $<..>$ means the average over many reference times as well as independent realizations.  The $C(t)$ is calculated for different size ratios ($S =8$, $10$, $12$), 
averaged over $160$ ensembles. 

\begin{figure*} [hbt]
 \centering    
 
 \includegraphics[width=0.85\textwidth]{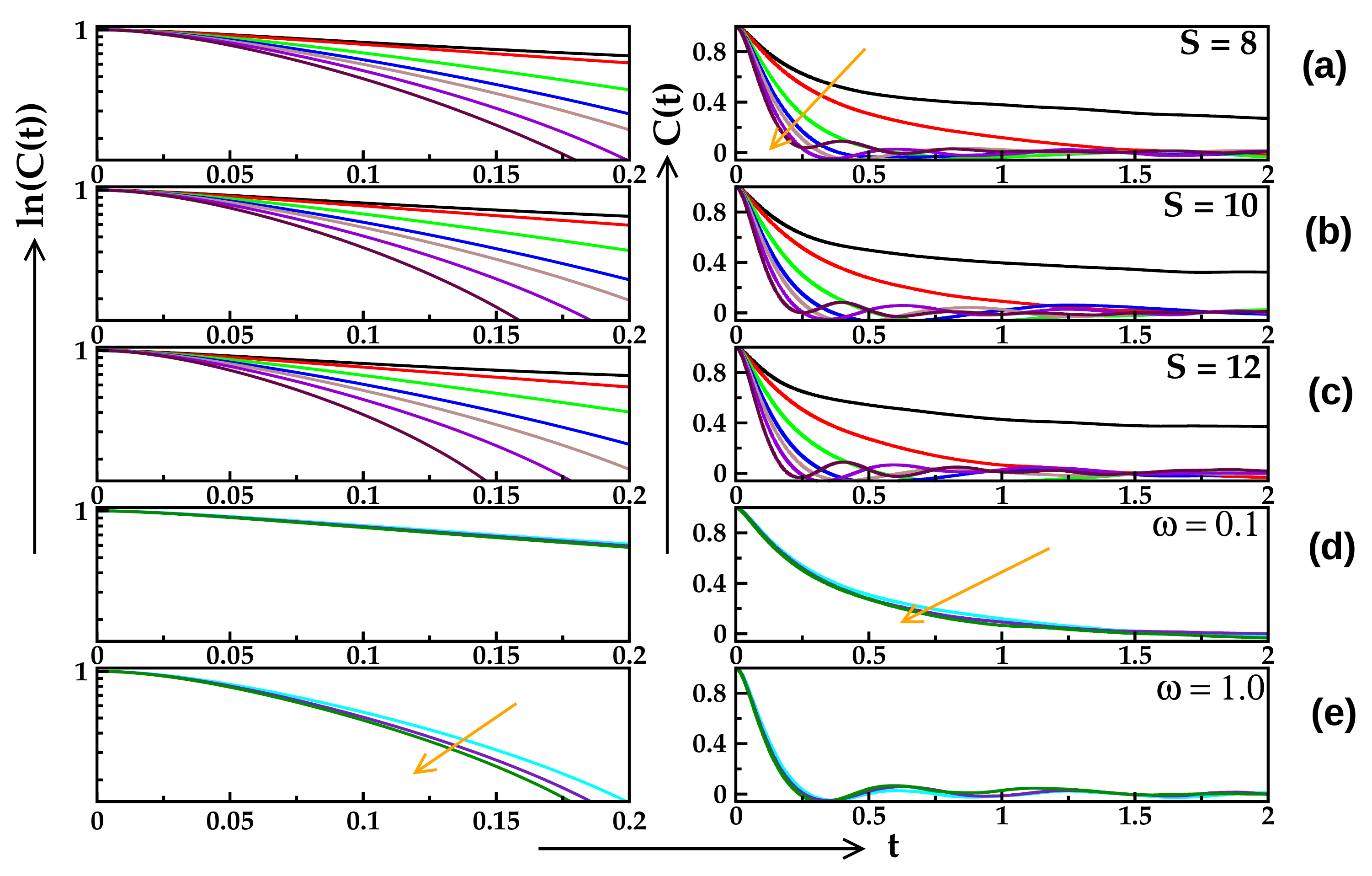}
 \caption{Orientation autocorrelation function 
C(t) of the torquer for different chirality values ($\omega=0,0.1,0.3,0.5,0.7,1.0,1.5$.) and size ratios ($S$ = 8,10, and 12). The right panel shows the plots in semi-log-y scale whereas the left panels are in linear scale. Panels (a)-(c) show that as chirality increases(along the direction of the arrow) C(t) decays more rapidly and develops oscillations
Panels (d) and (e) compare C(t) at chirality values, $\omega = 0.1$ and $1.0$ for two size ratios, $S$ = 8, 10, and 12(in the increasing order along the direction of the arrow). The semilog plot in panel (e) reveals a steeper initial decay for S = 12, which along with the linear plot in panel (e) shows that oscillations persist over longer durations for larger inclusions. These trends highlight the dependence of autocorrelation characteristics on the torquer size at fixed chirality.
}
 
 \label{fig.5}
 \end{figure*}

At low $\omega$, the autocorrelation decays slowly, indicating that the torquer undergoes weak and slower reorientation over time and retains angular memory over a large time. With increasing chirality, the autocorrelation decays more rapidly. For  $\omega > 0.3 $, distinct oscillations emerge in $C(t)$, signaling that the inclusion is rotating frequently and periodically. Each cycle of the oscillation corresponds approximately to a full angular revolution, with maxima and minima denoting recurrent alignment and anti-alignment with the initial orientation, respectively. In Fig. \ref{fig.5}(a-c), left panel shows the semi-log-y plot of $C(t)$  $vs$. time t, which shows that for all chiralities and size ratios the early time decay of $C(t)$ is exponential and becomes periodic at larger times. The periodicity is clear in right panel plot of $C(t)$ $vs$. time on linear scale. The periodic oscillations are clear for larger chiralities. In bottom two rows Figs. \ref{fig.5}(d-e) we show the plot of $C(t)$ $vs$. time for two $\omega = 0.1$ and $1.0$ respectively and for different size ratios. For smaller $\omega=0.1$, the $C(t)$ remains invariant by increasing $S$, whereas decay sharpens for larger $S$ at high $\omega = 1.0$, although the dependence on $S$ is much weaker than that on $\omega$.\\   
The overall behavior of $C(t)$ thus reveals a transition from slow angular dynamics at low chirality to coherent, persistent rotation at higher chirality. 
 \begin{figure*}[htb]
 \centering    
 \includegraphics[width=0.80\textwidth]{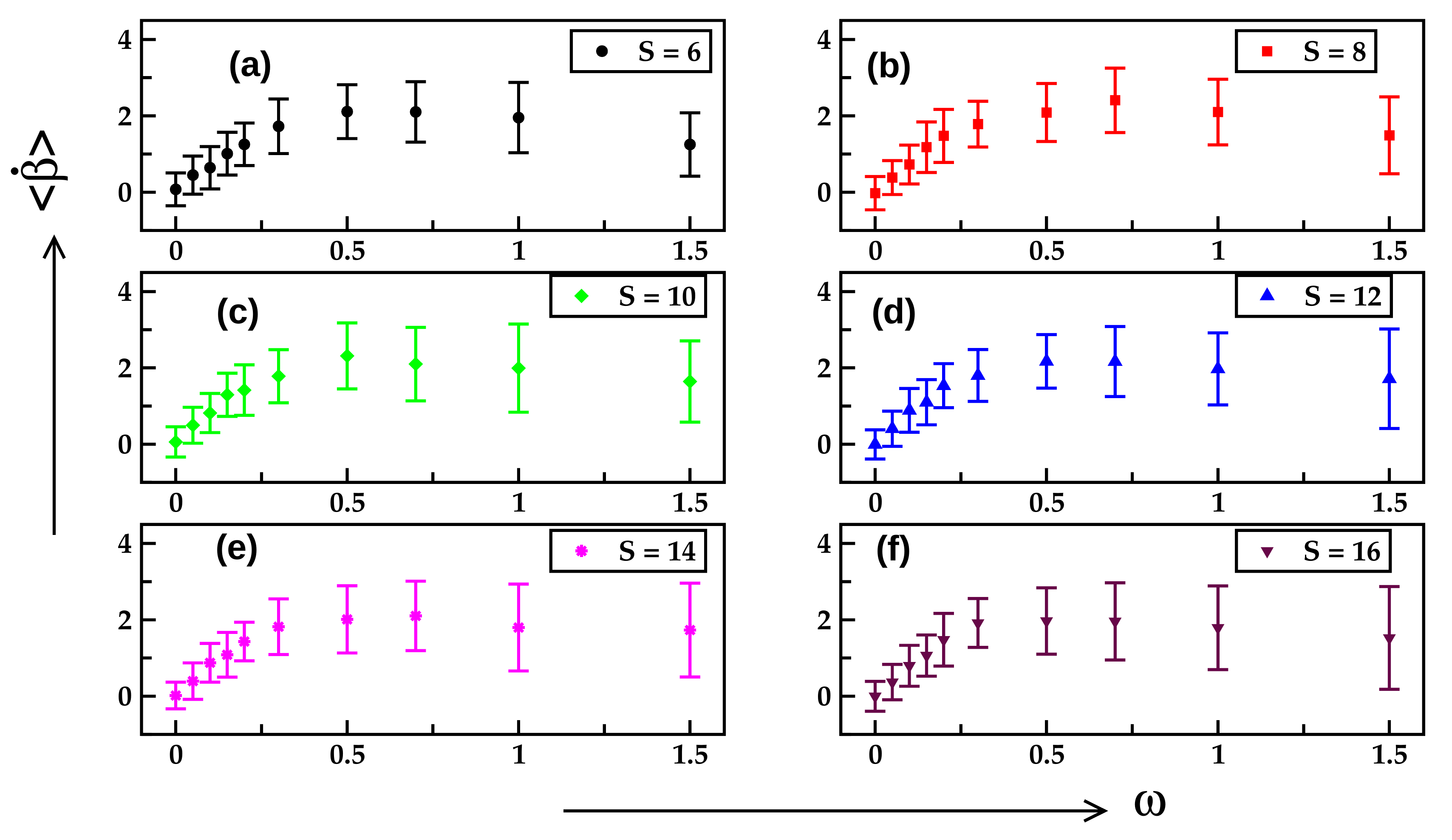}
 \caption{Fig.(a)-(e) shows mean angular velocity $\langle \dot{\beta} \rangle$ of the torquer as a function of chirality $\omega$ for size ratios $S = 6, 8, 10, 12, 14, 16$.} 
 
 \label{fig.6}
 \end{figure*}
In the higher chiral regime, active particles follow more curved trajectories due to stronger rotational drive, leading to more frequent re-encounters with the inclusion boundary. These interactions are shaped by the curvature of their paths and the constraint of volume exclusion, resulting in angularly biased contact patterns around the inclusion. While individual contacts may be brief, the consistent directional tendency of these interactions gives rise to a net torque, conducive to persistent rotation. \\

Now we characterize the magnitude of rotation of the inclusion by measuring its mean angular velocity. 
\begin{equation*}
\langle \dot{\beta} \rangle = \left\langle \frac{\beta(t + \Delta t) - \beta(t)}{\Delta t} \right\rangle_t.
\end{equation*}
Fig. \ref{fig.6} shows the variation of $\langle\dot{\beta}\rangle$ as a function of chirality $\omega$ for different size ratios $S$. $<..>$ has the same meaning as defined before. 
  At low $\omega$, the mean angular velocity is small, indicating weak or incoherent rotation. As chirality increases, the angular velocity rises sharply and reaches a peak at an intermediate value of $\omega$.
This trend implies that increased chirality initially enhances rotation, likely due to the emergence of more directionally consistent particle interactions. However, beyond the peak, the angular velocity does not continue to grow despite increased directional bias in the particle motion. Rather, it leads to a decline in $\langle\dot{\beta}\rangle$. This decline is more pronounced for smaller size ratios (Fig. \ref{fig.6}    (a,b)). In such cases, the inclusion's small surface cannot sustain the persistent rotation of active particles once chirality becomes too high. The rapidly turning active particles fail to establish or maintain persistent directional forces necessary for sustained rotation, resulting in a gradual drop in $\langle\dot{\beta}\rangle$. On the other hand, large inclusion (Fig. \ref{fig.6} (c-f)), with extended perimeters, continues to interact with a broader distribution of active particles, enabling a more sustained torque input even at higher chirality. This results in a slower, more gradual decline in $\langle\dot{\beta}\rangle$ for larger size ratios. This indicates that larger torquers are able to sustain faster rotation over a broader range of chirality values, likely due to greater opportunity for interaction with more active particles along their extended boundary.\\

\begin{figure*}[hbt]
 \centering    
 \includegraphics[width=0.85\textwidth]{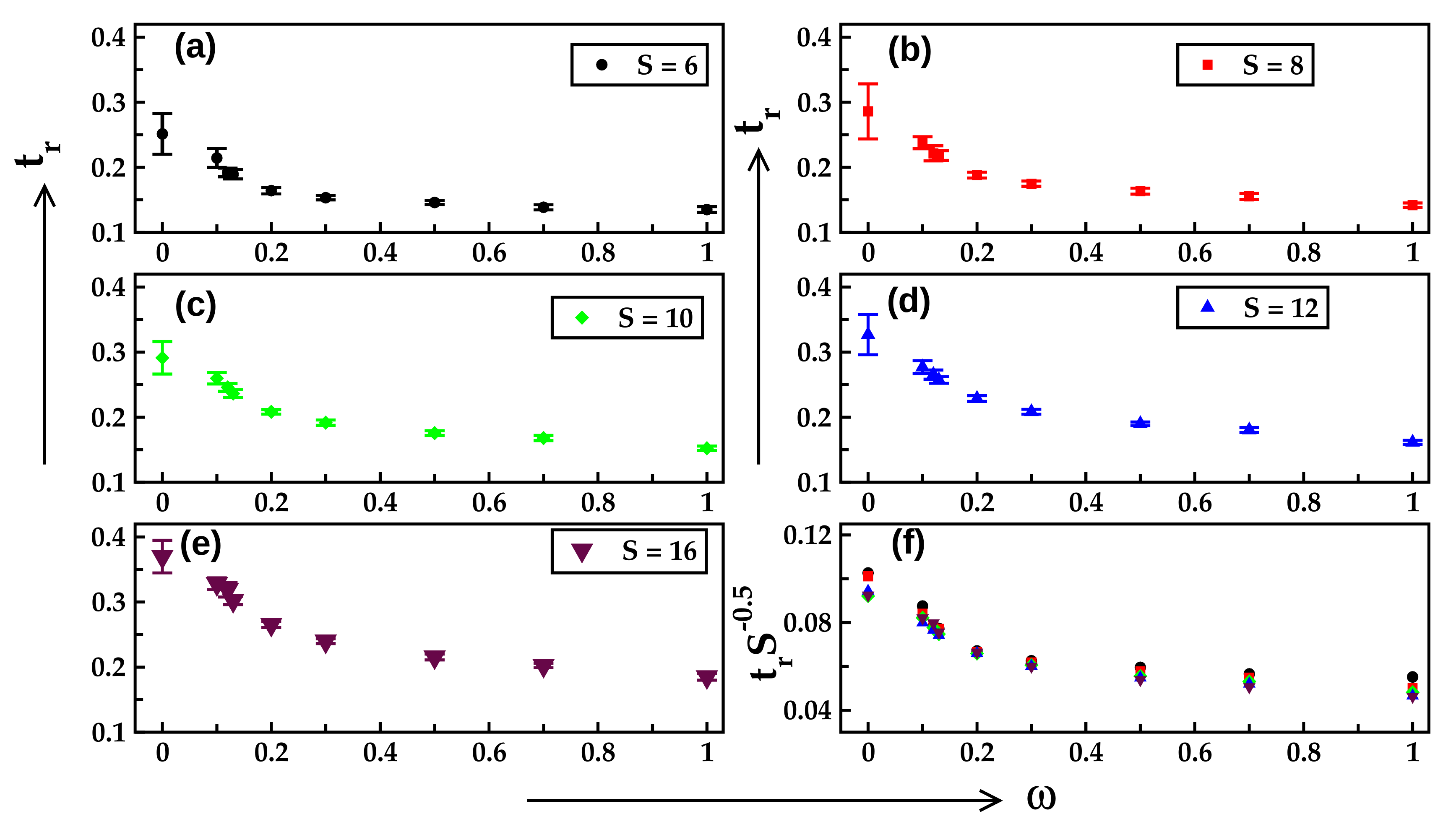}
 \caption{Fig.(a-e) shows the mean residence time $t_r$ of active particles near the torquer, as a function of chirality $\omega$ and for  size ratios
$S = 6, 8, 10, 12, 16$. Fig. (f) shows that the $t_r$ $vs$ $\omega$ plot for different $S$ values collapse to a single curve for scaled residence time $t_r/\sqrt{S}$.}
\label{fig.7}
 \end{figure*}

We further estimate how  much the interaction time of active particles  with the inclusion is influenced by their chirality. We then measure the mean residence time of a single active particle at the surface of the inclusion, where mean is calculated over all the active particles coming in contact with the inclusion's surface 
all of which stayed longer than or equal to $10^{-2} \tau$. Also mean is calculated over different independent realisations. In Fig. \ref{fig.7} (a-e) we show the variation of residence time $t_r$ (shown in units of $\tau$) $vs$. $\omega$ for different $S$. 
For all size ratios the $t_r$ decreases with increasing $\omega$, which is just due to the faster reorientation events of active particles on increasing $\omega$. For small $\omega \leq 0.1$ decay is slower and then shows a sharp decay for intermediate $0.1< \omega <0.2$ and for larger $\omega$ it flattens. Further in Fig. \ref{fig.7} (f) we show the scaled residence time $t_r$ $vs$. $\omega$ and find that data for all $S$ values collapse to a single curve with scaled $t_r /\sqrt{S}$. 

{\section{Discussion}\label{sec:dis}}
Our study demonstrates that persistent rotational dynamics can emerge in a geometrically symmetric inclusion solely due to its interaction with a bath of chiral active Brownian particles (ABPs). This result contrasts with earlier works that primarily attribute rotation due to shape asymmetry in the inclusion. By carefully tuning the chirality of the active particles and the size ratio between active particles and torquer, we uncover two distinct mechanisms that contribute to the angular motion of the torquer: torque imbalance due to spatial inhomogeneities and residence-time–driven torque generation.
At low chirality, rotational diffusion of active particles leads to persistent local clustering and density imbalance around the torquer, generating a net torque through uneven force contributions. These density inhomogeneities drives a slower but irregular rotation. The resulting angular dynamics is slow and lacks persistence, as evidenced by the gradual decay in orientation autocorrelation and the high correlation between torquer orientation and gap angle. This mechanism is stochastic and indirect, relying on spontaneous clustering and local jamming near the inclusion's perimeter.
In contrast, in the high-chirality regime, active particles exhibit persistent circular motion, with a strong bias in orientation dynamics. While this chirality enhances the coherence of the torque direction, it simultaneously reduces the residence time of active particles on the inclusion surface. The decreased contact duration limits the efficacy of torque transmission, causing a saturation or even decline in the angular velocity of the inclusion.\\
Our measurements of residence time directly support this interpretation: the peak in angular velocity corresponds to the crossover point where the directional coherence of particle motion outweighs the loss of contact time, beyond which the torque efficiency diminishes.
The non-monotonic dependence of inclusion's rotation on chirality thus emerges from a competition between directional order and interaction duration. Interestingly, increasing the size ratio $S$ enhances both mechanisms: it promotes stronger density fluctuations at low chirality and sustains longer contact arcs at high chirality. As a result, larger inclusion sustains higher angular velocities over broader chiralities, suggesting a potential control knob for optimizing rotational behavior.\\
These findings have broader implications for the design of synthetic micromachines and activity driven transport mechanisms. In particular, they demonstrate that shape asymmetry is not a necessary condition for generating persistent motion in active environments. Instead, chirality-induced alignment interactions alone can suffice. \\
\section{Conflicts of Interest}
There are no conflicts of interest to declare.
\section{Acknowledgement}
AP thanks the support and the resources provided by PARAM
Shivay Facility under the National Supercomputing Mision, Government of India at the Indian Institute of
Technology, Varanasi. SM thanks DSTSERB India, ECR/2017/000659, CRG/2021/006945 and
MTR/2021/000438 for financial support.

{\section{Appendix A : Description of animation}\label{sec:appndx}}
\textbf{Mov1}: The animation shows a single rotation  for size ratio $S = 8$ and $\omega = 0.5$.\\
\textit{Link}:\url{https://drive.google.com/file/d/1UyPH4yWlyh3bNhSezKwFXkKCkb8ScEj3/view?usp=sharing}.\\

\textbf{Mov2}: The animation shows a single rotation  for size ratio $S = 8$ and $\omega = 0.7$.\\
\textit{Link}:\url{https://drive.google.com/file/d/1DuxFXkHIc_7xHpGGTsRdoOlyZB2ppPxx/view?usp=sharing}.\\

\textbf{Mov3}: The animation shows a single rotation  for size ratio $S = 8$ and $\omega = 1.0$.\\
\textit{Link}:\url{https://drive.google.com/file/d/1O16_LyLRgO2_kL-PonSixeS5sZw7LEWd/view?usp=sharing}.\\

\bibliography{citation}

\end{document}